\documentclass[amssymb,aps,twocolumn,showpacs, superscriptaddress]{revtex4}
\usepackage{graphicx}
\usepackage{amsmath}
\usepackage{hyperref}
\usepackage{color}

\def\cO{\mathcal{O}}

\newcommand{\ket}[1]{\left| #1\right\rangle}        
\newcommand{\bra}[1]{\left\langle #1\right|}        

\begin{document}

\title{Quantum Speedup by Quantum Annealing}

\author{Daniel Nagaj}
\affiliation{Research Center for Quantum Information, Slovak Academy of Sciences, Bratislava, Slovakia}

\author{Rolando  D. Somma}
\affiliation{Los Alamos National Laboratory, Los Alamos, NM 87545, USA}
\author{M\'aria Kieferov\'a}
\affiliation{Research Center for Quantum Information, Slovak Academy of Sciences, Bratislava, Slovakia}

\date{\today}

\begin{abstract}
We study the glued-trees problem
 of Childs, et. al.~\cite{Childs:2002} in the adiabatic model of quantum
computing and provide an annealing schedule to solve an oracular problem exponentially faster than classically possible.
The Hamiltonians involved in the quantum annealing  do not suffer from the so-called sign problem.  Unlike the typical scenario,
our schedule is efficient even though the minimum energy gap of the Hamiltonians  is exponentially small in the problem size. We discuss generalizations based on initial-state randomization
to avoid some slowdowns in adiabatic quantum computing due to small  gaps.
\end{abstract}

\pacs{03.67.Ac, 03.67.Lx, 03.65.-w, 42.50.Lc}

\maketitle

Quantum annealing is a powerful
heuristic to solve problems in optimization~\cite{QA-together, Nishimori1998}. In quantum computing, the  method consists of preparing a 
low-energy or ground state $\ket \psi$ of a quantum system such that, after a simple measurement,
the optimal solution is obtained with large probability.  $\ket \psi$  is
prepared by following a particular annealing schedule, with a parametrized Hamiltonian
path subject to initial and final conditions. 
A ground state of the initial Hamiltonian is then transformed to $\ket \psi$ 
 by varying the parameter adiabatically.
 In contrast to more general quantum adiabatic state transformations,
 the Hamiltonians along the path in quantum annealing are termed {\em stoquastic}
 and
do not suffer from the so-called {\em numerical sign problem}~\cite{loh1990}:
for a specified basis, the off-diagonal
Hamiltonian-matrix entries are nonpositive~\cite{bravyi_stoquastic_2008}.
This property is useful for classical simulations~\cite{Nishimori1998}.

A sufficient condition for convergence of the quantum method
is given by the quantum adiabatic approximation.
It asserts that, if the rate of change of the Hamiltonian scales with
the energy gap $\Delta$ between their two lowest-energy states, $\ket \psi$ can be prepared with controlled accuracy~\cite{AA-together,Lidar-adiabatic-error}.
Such an approximation may  also be necessary~\cite{boixo:qc2009b}. However, it  could result in undesired overheads  if $\Delta$ is small but  transitions between the  lowest-energy states are forbidden due to selection rules, or
if transitions between lowest-energy states can be exploited to prepare $\ket \psi$. 
The latter case corresponds to the annealing schedule  in this Letter.
It turns out that the {\em relevant} energy gap for the adiabatic approximation in these cases
is not $\Delta$ and can be much bigger.

Because of the properties of the Hamiltonians,
the annealing can also be simulated using
probabilistic classical methods such as quantum
Monte-Carlo (QMC)~\cite{Nightingale1999}. The goal in QMC is to sample according to the distribution of the ground state, i.e. with probabilities coming from
  amplitudes squared.
 While we lack of necessary conditions
that guarantee convergence, the power of QMC
is widely recognized~\cite{Nightingale1999,Nishimori1998,Santoro2002}.
In fact, if the Hamiltonians satisfy an additional frustration-free property, efficient QMC simulations for quantum annealing exist~\cite{somma_2007,bravyi_2009}. 
This places a doubt on whether a quantum-computer
simulation of general quantum annealing processes
can ever be done using substantially less resources than QMC or
any other classical simulation.

Towards answering this question, we provide an oracular problem and 
give a quantum-annealing schedule that, on a quantum computer,
prepares a quantum state $\ket \psi$ encoding the solution.
The time required to prepare $\ket \psi$ is polynomial in the problem size, herein ${\rm poly}(n)$.
The oracular problem was first introduced in Ref.~\cite{Childs:2002} in the context of quantum walks,
where it was also shown that no classical method can give the solution
using ${\rm poly}(n)$ number of oracle calls. Our result thus places limits
on the power of classical methods that simulate quantum annealing,
even when the sign-problem is not present.

We remark that
the general question of existence of efficient classical simulations when $\Delta$ is $1/{\rm poly}(n)$
is not answered in this Letter. The annealing schedule we provide
is not intended to follow the ground state in the path; transitions 
to the closest (first-excited) eigenstate are allowed. Nevertheless, the system (almost) remains in the subspace spanned by these two states at all times. 
There are regions in the path where  $\Delta \propto \exp(-n)$. 
We  induce transitions in that subspace by choosing an annealing rate that
is much larger than $\Delta$, i.e. at $1/{\rm poly}(n)$ rates. Contrary to the typical case, such transitions are useful here. They guarantee that $\ket \psi$
is prepared after the annealing due to a symmetry argument:
The same type of transition that transforms the ground to the first-excited state, later transforms
the first-excited state back to the final ground state $\ket \psi$.

\begin{figure}
\begin{center}
\includegraphics[width=2.7in]{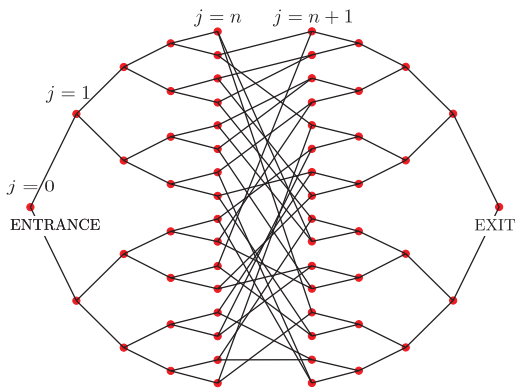}
\end{center}
\caption{Two binary trees of depth $n=4$ glued randomly. The number of vertices is $N=2^{n+2}-2$.
Each vertex is labeled with a randomly chosen $2n$-bit string. $j$ is the column number.}
\label{glued-trees}
\end{figure}

In more detail, we consider the oracular problem from Ref.~\cite{Childs:2002} that 
is defined as follows. 
We are given an oracle that consists of the adjacency
matrix $A$ of two binary trees that are randomly {\em glued} (by a random cycle) as in Fig~\ref{glued-trees}.  
Specifically, there are $N \in \cO(2^n)$ vertices named with randomly chosen $2n$-bit strings.
The oracle outputs the names of the adjacent vertices on any given input vertex name.
There are two special vertices, ENTRANCE and EXIT -- the roots of the 
binary trees. They can be identified because they are the only vertices of degree two
in the graph. The problem is: Given an oracle $A$ for the graph and the name of the ENTRANCE,
find the name of the EXIT. As mentioned, no classical algorithm can output the name of the EXIT,
with high probability, using less than a subexponential (in $n$) number of oracles. A quantum walk-based
algorithm can solve this problem efficiently with bounded probability~\cite{Childs:2002}. However, a quantum annealing
method to solve this problem efficiently using simple (stoquastic) Hamiltonians remained unknown. 
In this Letter, we present a new quantum annealing approach for this problem that  provides the solution with arbitrarily high probability, c.f. Refs.~\cite{Lidar-adiabatic-error,Wiebe-adiabatic-error}.

We assume a Hamiltonian version of the oracle
so that evolutions under $A$ can be implemented.
We also allow for evolutions under $H_0$ and $H_1$, these being the diagonal
Hamiltonians that distinguish the ENTRANCE and EXIT, respectively. 
Such evolutions can  be realized efficiently~\cite{aharonov_adiabatic_2003,cleve_quantum_2009}, i.e.
using $\cO(T)$ oracles for evolution time $T>0$.
We let $a(V) \in \{0,1\}^{2n}$ be the name of vertex $V$. Then, $H_0 \ket {a'} =- \delta_{a({\rm ENTRANCE}),a'}
\ket {a'}$ and $H_1 \ket {a'} =- \delta_{a({\rm EXIT}),a'}
\ket {a'}$, so that their ground states encode $a({\rm ENTRANCE})$ and $a({\rm EXIT})$,
respectively. 
The Hamiltonian path for the annealing  will consist of a specific interpolation involving
$H_0$, $A$, and $H_1$.

As in Ref.~\cite{Childs:2002}, we find it  useful to define the (orthonormal)  states 
\begin{align}
\ket{{\rm col}_j} = \frac 1 {\sqrt{N_j}} \sum_{i \in j \text{th column}} \ket {a(i)} \; . \label{columns}
\end{align}
These are uniform-superposition states over all states labeled by the
names of vertices at the $j-$th column.
$N_j = 2^j$ for $0 \le j \le n$ and $N_j=2^{2n+1-j}$ for $n+1 \le j \le 2n+1$; see Fig.~\ref{glued-trees}. In particular, $\ket{{\rm col}_0} =\ket{a({\rm ENTRANCE})} $ and 
$\ket{{\rm col}_{2n+1}} =\ket{a({\rm EXIT})} $.
We observe that the subspace spanned by $\{\ket{{\rm col}_j} \}_{0 \le j \le 2n+1}$ is invariant under the action of  $A$, $H_0$, and  $H_1$. In the  basis determined by Eqs.~(\ref{columns}), $A$ has non-zero matrix elements  in its first off-diagonals only. For simplicity, we redefine $A \gets \sqrt 2 A$ so that the matrix elements are
\begin{align}
\label{eq:Amatrix}
\bra{{\rm col}_j} A \ket{{\rm col}_{j+1}} = \left \{ \begin{matrix} \sqrt 2 & j=n \cr 1  & {\rm otherwise.} \end{matrix} \right. 
\end{align}
Also,
\begin{eqnarray} 
\bra{{\rm col}_j} H_0 \ket{{\rm col}_{j}} &=& - \delta_{j,0} \; ,\\
\bra{{\rm col}_j} H_1 \ket{{\rm col}_{j}} &=& -\delta_{j,2n+1} \; . \nonumber
\end{eqnarray}

We choose the Hamiltonian path 
\begin{eqnarray}
   H(s) = (1-s) \alpha H_0 - s(1-s) A + s \alpha H_1 \label{ourH}
\end{eqnarray}
that interpolates between $H_0$ and $H_1$ for  $0 \le s \le 1$. The parameter  $\alpha$ is independent of $n$ and satisfies
$0<\alpha<1/2$.
We will show that using the Hamiltonian path of Eq.~\eqref{ourH}, and annealing
at a rate $\dot s(t) \propto 1/{\rm poly}(n)$, the resulting evolution
transforms $\ket{a({\rm ENTRANCE})}$ to a state that has arbitrarily
high overlap with  $\ket{a({\rm EXIT})}$.

\begin{figure}
\begin{center}
\includegraphics[width=3.4in]{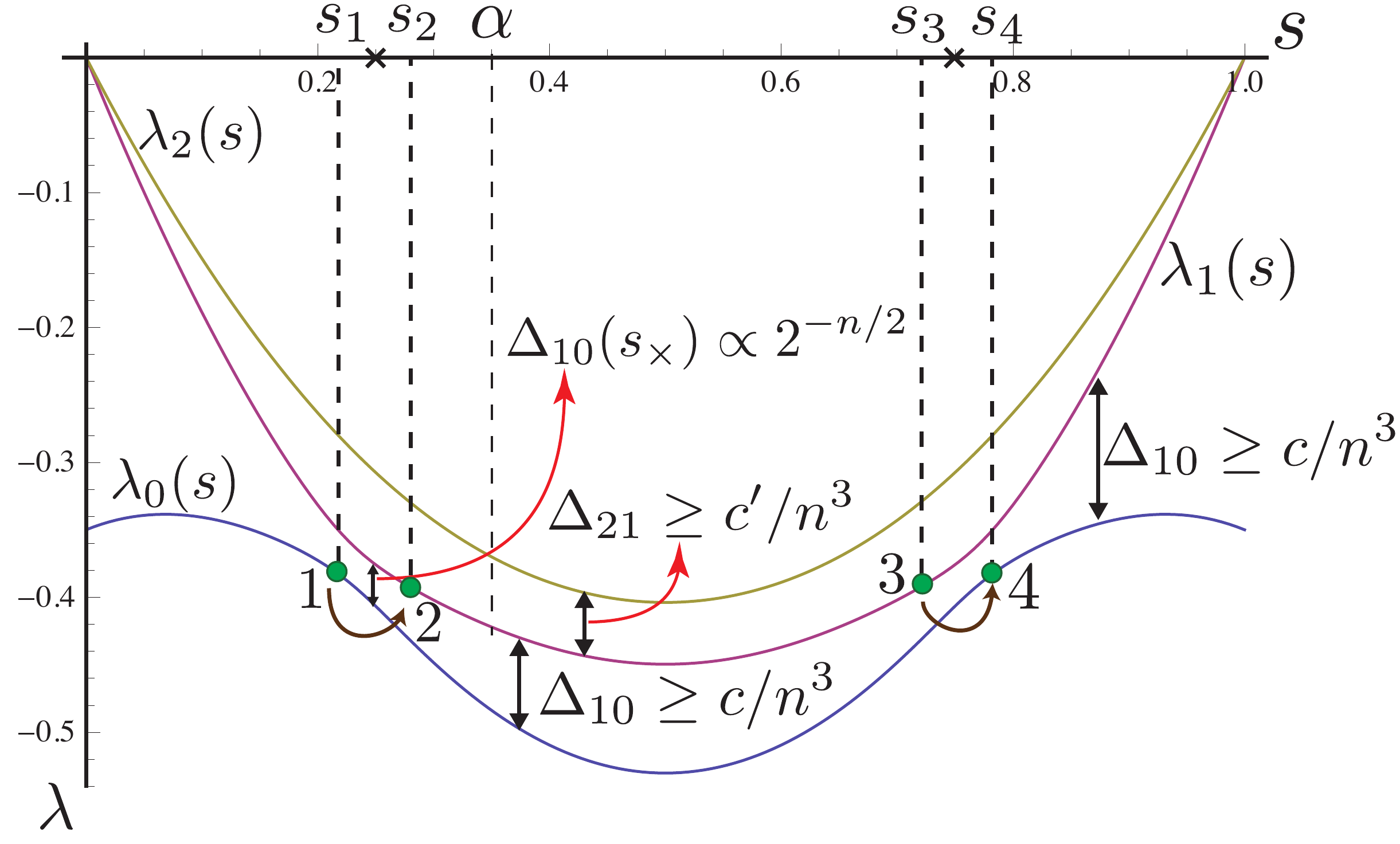}
\end{center}
\caption{The three lowest eigenvalues of $H(s)$ in the subspace spanned by the
states of Eqs.~\eqref{columns}, for $\alpha=1/\sqrt 8$ and $n=10$.
$\Delta_{jk}=\lambda_j -\lambda_k$ 
is the  gap between  $\lambda_j$ and $\lambda_k$,
the eigenvalues corresponding to the $j$-th and $k$-th excited states, respectively. 
We divide the evolution in five stages according to $s_1$, $s_2$, $s_3$, and $s_4$ (see text),
with $s_1 < s_\times = \alpha/\sqrt2 = 0.25< s_2$ and $s_3=1-s_2< 1-s_\times < s_4=1-s_1$.
Inside $[s_1,s_2]$ and $[s_3,s_4]$, the gap $\Delta_{10}(s)$ becomes
exponentially small in $n$. Elsewhere, $\Delta_{10}$ is only polynomially small in $n$.
Brown arrows depict level transitions
for an annealing rate in which $\dot s(t) \propto 1/{\rm poly}(n)$. Other gap scalings
are also shown.}
\label{eigenvalues}
\end{figure}

\noindent{\em Spectral properties}--- 
To prove the efficiency of the quantum method we utilize the spectral properties of $H(s)$;
particularly relevant are the spectral gaps.
The following analysis is valid if we restrict to the invariant subspace spanned by  Eqs.~\eqref{columns}. 
Figure~\ref{eigenvalues} shows the
three lowest eigenvalues of $H(s)$, obtained numerically, in this subspace.
This suggests a particular eigenvalue behavior.
We can analytically study the Hamiltonians 
by proposing the {\em ansatz} $\ket{\phi}= \sum_j \gamma_j \ket{{\rm col}_j} $,
with 
\begin{align}
\label{eq:ansatz}
\gamma_j &= a e^{i p j} + b e^{-i p j} \; , \;  0 \le j \le n \; , \\
\nonumber
\gamma_j &= c e^{i p (2n+1-j)} + d e^{-i p(2n+1-j)} \; , \;  n+1 \le j \le 2n+1 \; ,
\end{align}
and $p \in \mathbb{C}$.
The eigenvalue condition $H \ket \phi =\lambda \ket \phi$ and $\bra \phi \phi \rangle=1$
allow us to find
expressions for $a$, $b$, $c$, $d$, and $\lambda$. In particular, the eigenvalues are
$\lambda = -2s(1-s) \cos p$.
We provide a more detailed analysis of the spectrum in the supplemental
online material  (Supp. Mat.) and
present only the relevant results  here. Because of the $s \leftrightarrow (1-s)$ symmetry,
 it suffices
to analyze the parameter region $0 \le s \le 1/2$.

In the following, $x \approx_ \epsilon y$ if $|x-y |\le \epsilon$ and $\epsilon \in \cO(2^{-n/2})$.
When $n \rightarrow \infty$, the two lowest eigenvalues cross, or become equal, at $s=s_\times=\alpha/\sqrt 2$.
Different eigenvalue behavior is  obtained at both sides of $s_\times$.
For $n < \infty$ and $0 \le s \le s_\times$, the spectral gap between the two
lowest eigenvalues is
\begin{align}
\nonumber
\Delta_{10}(s) &= \lambda_1(s) - \lambda_0(s) \\
& \approx_\epsilon- (1-s) \left(  \frac {3s} {\sqrt 2} -  \frac {\alpha^2 + s^2} {\alpha} \right) \; .
\label{secondS0}
\end{align}
The eigenvalue crossing is avoided for $n<\infty$ but the spectral gap $\Delta_{10}(s)$ is
exponentially small in $n$ near $s_\times$. Also, $\Delta_{10}(s) \ge c/n^3$ for $0 \le s \le s_1=  s_\times - \delta$, with $\delta \in \Omega(1/n^3)$. 
Because the ordering of the two lowest eigenvalues
{\em swaps} for $ s > s_\times$, $\Delta_{10}(s) \ge c/n^3$ in the region $s_2=s_\times + \delta \le s \le 1/2$.

The second excited state has an eigenvalue that corresponds to $p \approx_\nu \pi/(n+1)$,
with $\nu \in \cO(1/(n+1)^2)$. The spectral gap with the first excited state for $0 \le s \le \alpha$
is
\begin{align}
\nonumber
\Delta_{21}(s)& = \lambda_2(s) - \lambda_1(s) \\
& \approx _{\epsilon/2 + \nu} - s(1-s) \left( \cos \left( \frac \pi {n+1}\right)-\frac 3 {\sqrt 2}\right) \; .
\end{align}
In particular, $\Delta_{21}(s) \in \Omega(1)$ in the region $s_1 \le s \le \alpha$.

For $\alpha \le s \le 1/2$, the second eigenvalue also corresponds to $p \approx_\nu \pi/(n+1)$.
To bound the gap with the third eigenvalue, a more detailed analysis that approximates $p$ at order $1/(n+1)^2$ is carried in the Supp. Mat.. It results in
$\Delta_{21}(s) \ge c'/ n^3$, for some $c' >0$  (see Fig.~\ref{eigenvalues}).

\noindent{\em Annealing schedules}--- 
We use the following adiabatic approximation from Refs.~\cite{AA-together}.
Let the initial state be an eigenstate of $H(s_0)$ 
and $\Delta(s)$  the spectral gap to the nearest (non-degenerate)
eigenstate in some region $s_0 \le s \le s_f$. Then, an annealing rate of 
 $\dot s(t) \propto \varepsilon \Delta^2(s)$ or smaller suffices to prepare
 the eigenstate of $H(s_f)$ at error amplitude smaller than $\sqrt{\varepsilon(s_f-s_0)}$; i.e. the 
 overlap between the evolved state and the eigenstate is at least $\sqrt{1-\varepsilon(s_f-s_0)}$.
(Better error scaling is possible~\cite{Lidar-adiabatic-error,Wiebe-adiabatic-error}.)
 
To prove that  $\dot s (t) \propto \varepsilon/n^6$ suffices
to transform $\ket{a({\rm ENTRANCE})}$ to $\ket{a({\rm EXIT})}$ with large overlap, we split the 
evolution according to $[0,1] = \bigcup_{i=1}^5 V_i$, with $V_1=[0,s_1)$, $V_2=[s_1,s_2)$, $V_3=[s_2,s_3)$,
$V_4=[s_3,s_4)$, and $V_5=[s_4,1]$. The values of $s_i$ were determined previously; see Fig.~\ref{eigenvalues}. We write $\ket{\phi_0(s)}$ and $\ket{\phi_1(s)}$ for the ground and first excited
states, respectively. Note that $\ket{\phi_0(0)}=\ket{a({\rm ENTRANCE})}$ and $\ket{\phi_0(1)}=\ket{a({\rm EXIT})}$.
Then, due to the gap bounds and the adiabatic approximation, the following transformations
occur:
\begin{align}
\label{eq:transfs1}
\ket{\phi_0(0)} &\rightarrow_{\sqrt{\varepsilon s_1}} \ket{\phi_0(s_1)} \; , \\
\nonumber
\ket{\phi_1(s_2)} &\rightarrow_{\sqrt{\varepsilon (s_3- s_2)}} \ket{\phi_1(s_3)} \; , \\
\nonumber
\ket{\phi_0(s_4)} &\rightarrow_{\sqrt{\varepsilon (1-s_4)}} \ket{\phi_0(1)} \; .
\end{align}
 $\rightarrow_x$ indicates that the transformation occurred at error amplitude of order $x$.

Because $\Delta_{21}(s) \in \Omega(1)$ for $s \in V_2$, 
transformations between the ground or first-excited state and the second
excited state occur with amplitude  smaller than $\sqrt \varepsilon$.
Thus, all relevant transitions in $V_2$  occur in the 
manifold spanned by $\{ \ket{\phi_0(s)} ,\ket{\phi_1(s)} \}$.
For our annealing rate,
the following transformations occur with large amplitude (see below):
\begin{align}
\label{eq:transfs2}
\ket{\phi_0(s_1)} &\rightarrow \ket{\phi_1(s_2)} \; , \\
\nonumber
\ket{\phi_1(s_3)} &\rightarrow \ket{\phi_0(s_4)} \; . 
\end{align}

To obtain the approximation errors for Eqs.~\eqref{eq:transfs2}, we introduce 
the state $\ket u$ that is a uniform superposition
over all vertex names:
\begin{align}
\ket u = \frac 1 {\sqrt N} \sum_{i \in {\rm graph}} \ket{a(i)} = \sum_{j=0}^{2n+1} \sqrt{\frac{N_j}{N}} 
\ket{{\rm col}_ j } \; .
\end{align}
Here, $N_j=2^j$ for $0 \le j \le n$  and $N_j = 2^{2n+1-j}$ for $n+1 \le j \le 2n+1$.
Interestingly, $\ket u$ is {\em almost} an eigenstate for all $s$:
$H(s) \ket u \approx_{\epsilon/2}  -(s(1-s)3/\sqrt 2) \ket u$ and $\epsilon \in \cO(2^{-n/2})$ (see Supp. Mat.).
We define $f(t) = | \bra u U(t) \ket u |^2$, where
$U(t)$ is the evolution operator and $f(0)=1$.
Schr\"odinger's equation yields
\begin{align}
\nonumber
\dot f(t) &= -i \bra u H(s(t)) U(t) \ket u \bra u U^\dagger(t) \ket u + c.c. \\
& \approx_\epsilon 0 \; .
\end{align}
If $T\in \cO(n^3)$ is the evolution time to change $s$ from $s_1$ to $s_2$
with our annealing schedule ($|s_2 - s_1| \in \Omega(1/n^3)$),
we have $f(T) \approx_{\epsilon'} 1$ for  $\epsilon' \in \cO(\epsilon n^3)$.
In addition (Supp. Mat.),
\begin{align}
|\bra u \phi_1(s_1) \rangle| \approx_{\epsilon'} 1 \; , \; |\bra u \phi_0(s_2) \rangle| \approx_{\epsilon'} 1 \; ,
\end{align}
resulting in
\begin{align}
|\bra{\phi_1(s_1)} U(T) \ket{\phi_0(s_2)}|^2 \approx_{5 \epsilon'} 1 \; .
\end{align}
The transformation $\ket{\phi_1(s_1)} \rightarrow_{\sqrt{5 \epsilon'}} \ket{\phi_0(s_2)}$
then occurs.
Moreover,
because $U(t)$ is unitary, we also have
$\ket{\phi_0(s_1)} \rightarrow_{\sqrt{5 \epsilon'}} \ket{\phi_1(s_2)}$,
and from
symmetry arguments, $\ket{\phi_1(s_3)}  \rightarrow_{\sqrt{5 \epsilon'}} \ket{\phi_0(s_4)}$. 
These level transitions are shown
in Fig.~\ref{eigenvalues}. Together with the transformations in Eqs.~(\ref{eq:transfs1}),
they prove the success of our quantum annealing method. Because $\epsilon' \ll 1$ 
for large $n$, the overall amplitude error is dominated by that of the adiabatic approximation.
This is of order $\sqrt{\varepsilon}$, with $\varepsilon$ arbitrary.

\begin{figure}
\begin{center}
\includegraphics[width=3in]{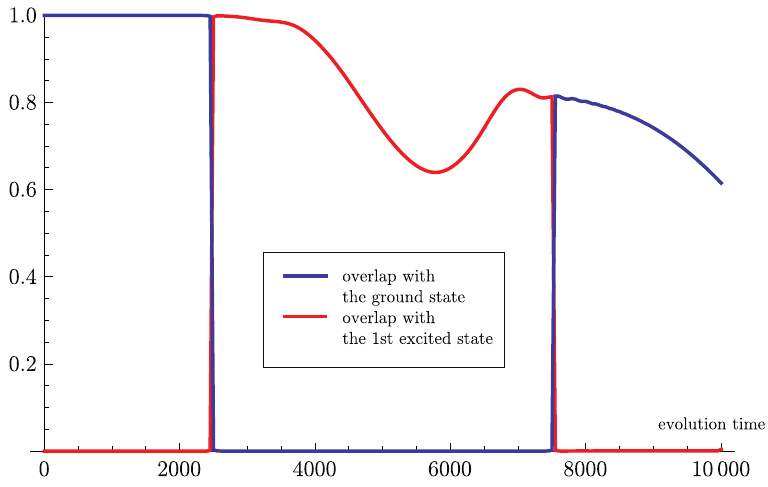}
\end{center}
\caption{Overlap of the evolved state $U(t) \ket{a({\rm ENTRANCE})}$
with the ground and first excited states as a function of time. 
In this case, $n=40$, $\alpha=1/\sqrt 8$, and $s(t)=t/10000$. The simulation
was performed in the column subspace.
}
\label{fig:overlaps}
\end{figure}

In Fig.~\ref{fig:overlaps} we show the overlaps of $U(t) \ket{\phi_0(0)}$
with the ground and first excited states
as a function of the evolution time, using our $\dot s(t)$, showing evidence for the transition among the
two low energy levels.

\noindent{\em Initial state randomization: a generalization}--- 
A method that guarantees a successful final state preparation
with probability $1/2$ in this case is possible if we randomize
the initial state preparation:  randomly prepare $\ket{a({\rm ENTRANCE})}$
or $\ket u$ that is almost   the first excited state. Both states
can be prepared efficiently. If we choose an annealing rate $\dot s(t) \propto \epsilon/n^6$
that forbids transitions to the second excited level, we are guaranteed, from unitarity,
that the final state is prepared with probability $1/2$. This method takes advantage of the fact
that the gaps with the third eigenvalue are of order $1/n^3$ in the region of interest. 

This randomized method can be carried to those cases
in which the $k$ lowest eigenvalues are at gaps $1/{\rm poly}(n)$ with the remaining ones.
Then, any eigenstate in this manifold can be prepared with probability
of order $1/k$ by randomizing the initial state preparation and choosing an adequate
annealing schedule. If $k$ is not too big, the method may be efficient regardless of 
whether the $k$ lowest eigenvalues have small gaps or cross. It assumes
the efficient preparation of any of the $k$ initial states.

General Hamiltonians do not satisfy
the assumptions; however, some Hamiltonians of physical systems
satisfy them. An example is the one-dimensional spin-1/2 Ising model in a constant transverse field: $H(J)
=J \sum_i \sigma_z^i \sigma_z^{i+1} + \sigma_x^i$, with $\sigma_\alpha^i$ being the Pauli operator
on spin $i$.
As the Ising coupling $J$ is changed from $0$ to $|J| \gg 1$, the two lowest eigenvalues
have a spectral gap that decreases exponentially with the system size $n$. However, the third
eigenvalue is always at a distance $1/{\rm poly}(n)$ from the two lowest ones.
This is a common property for  systems that have conformal invariance.


\medskip
\noindent{\em Conclusion}---
We provided an example of an oracular problem
for which a quantum adiabatic evolution can find 
the solution efficiently whereas exponential time 
is required for any classical method. The Hamiltonians
in the evolution do not suffer from the sign problem. 
Our result is a step towards {\em proving} the power
of quantum methods for quantum annealing.
We discussed the details of why our algorithm works
and how it can be generalized, under some assumptions
on the spectrum, by randomizing the initial state preparation.


\medskip
\noindent{\em Acknowledgements}--- We  thank S. Jordan, A. Childs, and A. Landahl for insightful discussions.
DN acknowledges support from the European project
Q-ESSENCE, the Slovak Research and Development Agency under the contract No. LPP-0430-09, and APVV COQI. RS acknowledges partial support from the National Science Foundation through the CCF program, and the Laboratory Directed Research and Development Program at Los Alamos National Laboratory
and Sandia National Laboratories.
 Sandia National Laboratories is a multi-program laboratory managed and operated by Sandia corporation, a wholly owner subsidiary of Lockheed Martin corporation, for the US DOE NNSA, under contract DE-AC04-94AL85000.


\newpage

\section{Supplementary Online Material} 

We analyze the spectrum of 
\begin{eqnarray}
	H(s) = \alpha (1-s) H_0 - s(1-s) A + \alpha s H_1
	\label{appendixH}
\end{eqnarray}
for $0 \le s \le 1$ and $\alpha=1/\sqrt8 \approx 0.35$ for simplicity (see below). The matrix elements of $A$
are in Eq.~(\ref{eq:Amatrix}).
We rescale the Hamiltonian and study $H'(s) = H(s)/(s(1-s))=\alpha' H_0 - A + \beta' H_1$, with $\alpha'=\alpha/s$ and $\beta' = \alpha/(1-s)$.
We let $\ket{\phi_p(s)}$ be an eigenstate and
propose the {\em ansatz} $\ket{\phi_p(s)}=\sum_{j=0}^{2n-1} \gamma_j \ket{{\rm col}_j}$ given by
\begin{align}
\gamma_j &= a e^{i p j} + b e^{-i p j} \; , \;  0 \le j \le n \; , \\
\nonumber
\gamma_j &= c e^{i p j'} + d e^{-i p j'} \; , \;  n+1 \le j \le 2n+1 \; ,
\end{align}
with $j' = 2n+1-j$. The constants $a,b,c,d$ are determined below. $p \in \mathbb{C}$
so that different solution behavior is obtained for real or imaginary $p$.

From the eigenvalue equation $H'(s) \ket{\phi_p(s)} =  \lambda'(s) \ket{\phi_p(s)}$,
we obtain 
\begin{align}
\lambda'& = -2 \cos p \; ,  
\end{align}
and the conditions 
\begin{align}
\label{eq:coeff1}
  \alpha' (a+b) & =  a e^{-ip} + b e^{ip} \; ,  \\
\nonumber
   \beta' (c+d) & =  c e^{-ip}  +d e^{ip}  \; , \\
   \label{eq:n,n+1}
a e^{ip(n+1)} + be^{-ip(n+1)} & = \sqrt 2 (c e^{ipn} + d e^{-ipn}) \; ,  \\
\nonumber
c e^{ip(n+1)} + de^{-ip(n+1)} & = \sqrt 2 (a e^{ipn} + b e^{-ipn})  \; . 
\end{align}
Together with the normalization condition $1=\bra{\phi} \phi\rangle$, these determine
the unknown parameters $p, \lambda, a, b , c ,d$. Simple inspection of Eqs. \eqref{eq:coeff1},\eqref{eq:n,n+1}
implies that $p$ must obey the quantization condition
\begin{eqnarray}
	f(p,n,\alpha') \, f(p,n,\beta') = 2,  
	\label{quantization1}
\end{eqnarray}
with
\begin{eqnarray}
	f(p,n,\theta) =   \frac{\sin ((n+2)p) - \theta \sin ((n+1)p)}{\sin ((n+1)p) - \theta \sin (np)} \; .
	\label{quantizationF}
\end{eqnarray}
The symmetry $\alpha' \leftrightarrow \beta'$ is evident in Eq.~\eqref{quantization1} and thus it suffices to consider
$\alpha',\beta'$ for which $0< s \le 1/2$. The nature of the solutions depends on whether
 $\alpha \geq 1/2$ or $\alpha<1/2$. 
Our choice of $\alpha=1/\sqrt 8<1/2$ will produce an efficient annealing method. In the following,
we use the condition of Eq.~\eqref{quantization1}
 to understand the behavior of the three lowest
 eigenvalues and to estimate the scaling of relevant energy gaps.

\subsubsection{Hyperbolic solutions}

We consider first the case where  $p=-iq$ is purely imaginary, 
so that $\sin p \rightarrow \sinh q$ and $\cos p \rightarrow \cosh q$
in Eq.~\eqref{quantizationF}. We  write $f_h(q,n,\theta)=f(-iq,n,\theta)$.
The behavior of the lowest eigenvalues for $n<\infty$
will be understood by considering the large-$n$ limit. 
In particular, $\lim_{n \rightarrow \infty} \sinh((n+1)\bar q_0) - \alpha' \sinh (n\bar q_0)=0$
for $\bar q_0 = \ln (\alpha')$. Also, $\lim_{n \rightarrow \infty} f_h (\bar q_0 , n , \beta')=\alpha'$
if $s \le \alpha<1/2$ (or $\alpha' \ge 1$), which implies $\beta' < \alpha'$.
Both limiting Eqs. imply that $\bar q_0$ determines an eigenvalue in the large-$n$
limit. Another eigenvalue can be obtained by noticing that
\begin{align}
\lim_{n \rightarrow \infty }f_h (\bar q_1,n , \alpha')f_h (\bar q_1,n , \beta')=2
\end{align}
if $\bar q_1 = \ln \sqrt 2$.
The eigenvalues of the original Hamiltonian of Eq.~\eqref{appendixH} are ($s<1/2$)
\begin{align}
\nonumber
F(s) &= -2 s(1-s) \cosh \bar q_0 = -(1-s) (\alpha + s^2/\alpha) \; ,\\
\nonumber
G(s) & =-2 s(1-s) \cosh \bar q_1 =-s(1-s) 3/\sqrt 2 \; , 
\end{align}
respectively. Note that $F(s ) = G(s)$ if $s=s_\times=\alpha/\sqrt 2$,
where a level crossing for $n \rightarrow \infty$ occurs.

The eigenvalues for $n<\infty$  are {\em slightly} different from those for $n < \infty$.
We are particularly interested in the energy gap around $s_\times$.
For $n=\infty$, a solution to Eq.~(\ref{quantization1}) is obtained if we let
 $\bar q = \ln \sqrt 2 +\epsilon (s)$ 
and do a second-order Taylor series approximation;
i.e. we approximate $2 \sinh (n \bar q) \approx (\sqrt 2)^n (1+n\epsilon(s) + n^2 \epsilon^2(s)/2 - 2^{-n})$.
Plugging the approximation in Eq.~(\ref{quantization1}), we obtain two solutions
with $\epsilon(s) \le \epsilon \in \cO(2^{-n/2})$. This finite size
correction also carries to the eigenvalues, proving an exponentially small gap of order $\epsilon$
near $s_\times$.
Away from $s_\times$, finite size corrections can be shown
to be of smaller order, if $s < \alpha$, by using a first-order Taylor series approximation.

We write $x \approx_\epsilon y$ if $|x-y|\le \epsilon$.
Then, for 
$s \le s_\times$, the lowest eigenvalue for $n <\infty$ is
\begin{align}
\lambda_0(s) \approx_{\epsilon/2} F(s)= -(1-s) (\alpha + s^2/\alpha) \; .
\end{align}
Similarly, the eigenvalue for the first excited state in that region 
is
\begin{align}
\lambda_1(s) \approx_{\epsilon/2} G(s)=-s (1-s) 3/\sqrt 2 \; .
\end{align}
These two eigenvalues are {\em swapped} for  $s \ge s_\times$
(a level crossing in $n \rightarrow \infty$).
It implies that the  lowest eigenvalue for $1/2 \ge s \ge s_\times$
is
\begin{align}
\lambda_0(s) & \approx_{\epsilon/2}  G(s) = -s (1-s) 3/\sqrt 2 \; . 
\end{align}
The  eigenvalue for the first excited state in $\alpha > s \ge s_\times$ is
\begin{align}
\lambda_1(s) & \approx_{\epsilon/2}  F(s)=-(1-s) (\alpha + s^2/\alpha) \; .
\end{align}

The solution corresponding to $\bar q = \ln \sqrt 2 +\epsilon(s) $
is valid for all $s$, i.e. it  is the first excited
state for $s < s_\times $ and  the ground
state for $s > s_\times$. Using Eqs.~(\ref{eq:coeff1}) and~(\ref{eq:n,n+1}),
the corresponding eigenstate $\ket{\phi_{\bar q}(s)}$ does not depend on $s$
if $n \rightarrow \infty$ (and thus $\epsilon \rightarrow 0$):
\begin{align}
\lim_{n \rightarrow \infty} \ket{\phi_{\bar q}(s)} =  \ket u \; .
\end{align}
$\ket u$ is the 
uniform superposition state over all vertex names, i.e.
\begin{align}
\ket u = \frac 1 {\sqrt N} \sum_{i \in {\rm graph}} \ket{a(i)} 
= \sum_{j=0}^{2n+1}\sqrt{\frac{N_j}{2^{n+2}-2} }\ket{{\rm col}_j} \; ,
\end{align}
with $N_j = 2^j$ for $0 \le j \le n$ and $N_j = 2^{2n+1-j}$ for $n+1 \le j \le 2n+1$.
Finite size corrections to $\ket{\phi_{\bar q}(s)}$ can be obtained
from Eqs.~(\ref{eq:coeff1}) and~(\ref{eq:n,n+1}).
By simple matrix multiplication
we obtain ($\alpha <1/2$)
\begin{align}
H(s) \ket u \approx_{ \epsilon/2} G(s) \ket u \; ,
\end{align}
showing that $\ket u$ is {\em almost} an eigenstate for all $s$.
Also,
\begin{align}
|\bra u \phi_{\bar q} (0)\rangle |=|\bra u \phi_{\bar q} (1) \rangle| \approx_{2\epsilon} 1 \; ,
\end{align}
which follows from the spectrum of $H_0$ and $H_1$, and the orthogonality condition.
For other values of $s$, there is an energy gap $\Delta(s)$ between 
$\ket{\phi_{\bar q} (s)}$ and any other eigenstate. Recall that the eigenvalue
is $\approx_{\epsilon/2} G(s)$.  With no loss of generality,
$\ket u = o \ket{\phi_{\bar q} (s)} + \sqrt {1-o^2} \ket{\perp}$, where $\bra{\perp}\phi_{\bar q} (s)\rangle=0$.
We rescale the Hamiltonian by $-G(s)$ so that $H(s) \ket u \approx_{ \epsilon/2} 0$
and $H(s) \ket{\phi_{\bar q} (s)} \approx_{\epsilon /2} 0$. This implies
that $\sqrt{1-o^2} \Delta(s)  \approx_\epsilon 0$ or, equivalently,
\begin{align}
| \bra u \phi_{\bar q}(s) \rangle | \approx_{\epsilon^2/\Delta^2(s)} 1 \; .
\end{align}

\subsubsection{Goniometric solutions}
For a full  characterization of the three lowest eigenvalues,
we investigate the second excited state for $0 \le s \le 1/2$
as well as the first excited state for $\alpha \le s \le 1/2$.
To this end, we consider the eigenvectors of the goniometric type, i.e. those for which $p \in \mathbb{R}$, and consider the large-$n$ limit first.
Equation~\eqref{quantizationF} is singular for $p \rightarrow \frac{k \pi}{n+1}$.
Finite size corrections can  be obtained by allowing $p(s) =  \frac{k \pi}{n+1} +\frac{\pi x(s)}{(n+1)^2}$
and using Eq.~\eqref{quantizationF} to solve for $x(s)$. By inspection,  solutions for $k=1$
correspond to the first and second excited eigenstates. We are ultimately
interested in spectral gaps as they guarantee the convergence of the quantum annealing method.
If $\lambda_2(s)$ is the eigenvalue of the second excited state, it follows that,
for $0 \le s \le \alpha$,
\begin{align}
\nonumber
\Delta_{21}(s) \approx_{\epsilon/2 + \nu} s(1-s)\left[ - 2 \cos \left( \frac{ \pi}{n+1} \right) + \frac 3 {\sqrt 2} \right] \; .
\end{align}
Here, $\Delta_{21}(s) = \lambda_2(s) - \lambda_1(s)$ and $\nu \in \cO(1/(n+1)^2)$. 

Also by inspection of Eq.~\eqref{quantizationF}, $\Delta_{21}(s)$ takes its minimum value at $s=1/2$ in the region $ \alpha \le s \le 1/2$. In this case $\alpha'=\beta'=2 \alpha$ and the solutions for $p$ at $s=1/2$ satisfy
\begin{align}
\label{eq:midquantization}
\frac{\sin((n+2)p) -2 \alpha \sin((n+1)p)} {\sin((n+1)p) -2 \alpha \sin( n p)} =\pm  \sqrt 2 \; .
\end{align}
This is satisfied if $p = \pi/(n+1)$ for $\alpha=1/\sqrt 8$, which corresponds to the eigenvalue of the second
excited state:
\begin{align}
\lambda_2(1/2) = -\frac 1 2  \cos \left( \frac {\pi}{n+1} \right) \; ,
\end{align}
or, equivalently, $x(1/2)=0$ for this solution.
The eigenvalue $\lambda_1(1/2)$ can be well approximated
if we let $\sin((n+1+r)p) \approx -\pi (r+x(s))/(n+1)$ in Eq.~\eqref{eq:midquantization} and using
the minus sign in its {\em rhs}. With this approximation, the equation becomes linear in $x(s)$
and we obtain $x(1/2) =-2.82$.
It means that
\begin{align}
\lambda_1(1/2) \approx_\mu - \frac 1 2 \cos \left[ \frac \pi {n+1} \left( 1 - \frac{2.82}{n+1} \right) \right] \; ,
\end{align}
with $\mu \in \cO(1/(n+1)^3)$.

Therefore, the spectral gap between the second and first excited states, for $\alpha \le s \le 1/2$,
satisfies
\begin{align}
\nonumber
\Delta_{21}(s) &\ge \lambda_2(1/2) - \lambda_1(1/2) \\
& \ge c'/n^3 \; ,
\end{align}
for a constant $c' \approx 2.82 \pi$.


\end{document}